\documentstyle[11pt,moriond,epsfig]{article}

\def\Journal#1#2#3#4{{#1} {\bf #2}, #3 (#4)}

\def\PLB{{\em Phys. Lett.}  B}

\def\PRD{{\em Phys. Rev.} D}

\def\be{\begin{equation}}
\def\ee{\end{equation}}
\def\bea{\begin{eqnarray}}
\def\eea{\end{eqnarray}}

\begin{document}
\vspace*{4cm}
\title{MEASUREMENT OF LIGHT-CONE WAVE FUNCTIONS BY DIFFRACTIVE
DISSOCIATION}

\author{ DANIEL ASHERY \footnote{Representing Fermilab E791 Collaboration} 
}

\address{School of Physics and Astronomy, Raymond and Beverly Sackler
Faculty of Exact Sciences\\
 Tel Aviv University, Israel}

\maketitle\abstracts{
Diffractive dissociation of particles can be used to study their
light-cone wave functions. Results from Fermilab experiment E791 for
diffractive dissociation of 500 GeV/c $\pi^-$ mesons into di-jets
show that the $|q\bar {q}\rangle $ light-cone asymptotic wave function
describes the data well for $Q^2 \sim 10 ~{\rm (GeV/c)^2}$ or more.
}

\section{Light-Cone Wave Functions}

The internal structure of hadrons and photons bears directly on the
fundamental interactions of quarks and gluons that create the hadronic
bound
state. It is also an essential ingredient in understanding the hadronic
strong, electromagnetic and weak interactions. A very powerful description
of the hadronic structure is obtained through the light-cone wave
functions.
These are frame-independent and process-independent quantum-mechanical
descriptions at the amplitude level. They encode all possible quark and   
gluon
momentum, helicity and flavor correlations in the hadron. The light-cone
wave functions are constructed from the QCD light-cone Hamiltonian:   
$H_{LC}^{QCD} = P^+P^- - P^2_{\perp}$ where $P^{\pm} = P^0 \pm P^z$
\cite{bps}. The wave function $\psi_h$ for a hadron $h$ with mass $M_h$
satisfies the relation: $H_{LC}^{QCD}|\psi_h\rangle = M_h^2
|\psi_h\rangle$.\\

The light-cone wave functions are expanded in terms of a complete basis of 
Fock states having increasing complexity \cite{bps}. The negative 
pion has the Fock expansion:

\begin{eqnarray}
|\psi_{\pi^-}>&=& \sum _n <n|\pi^-> |n> \nonumber \\
     &=&\psi^{(\Lambda)}_{d\bar{u}/\pi}(u_i,\vec{k}_{\perp i})|\bar{u}d>\\
     & &+\psi^{(\Lambda)}_{d\bar{u}g/\pi}(u_i,\vec{k}_{\perp i})|\bar{u}dg>
        + \cdot \cdot \cdot \nonumber
\label{pifock}
\end{eqnarray}
representing expansion of the exact QCD eigenstate at scale $\Lambda$ 
in terms of non-interacting quarks and gluons.
They have longitudinal light-cone momentum fractions:
\begin{equation}
  u_i = \frac {k_i^+}{p^+} = \frac {k_i^0+k_i^z}{p^0+p^z}, \;\;\;\;\;\;
  \sum_{i=1}^n u_i =1
  \label{eq:xlc}
\end{equation}
and relative transverse momenta
\begin{equation}
  \label{eq:kplc}
\vec{k}_{\perp i} \;\;, \;\;\;\;\;\;
\sum_{i=1}^n \vec{k}_{\perp i} = \vec 0_{\perp}.
\end{equation}

\noindent
The first term in the expansion is referred to as the valence Fock state,
as it relates to the hadronic description in the constituent quark model.
Heavy states of non-relativistic quarks have minimal gluon content.  
This state must therefore have a very small size and hence a large mass.
The higher terms are related to the sea components of the hadronic structure,
are larger and lighter. It has been shown that once the valence Fock state
is determined, it is possible to build the rest of the light-cone wave
function \cite{anto,dal}.

The essential part of the wave function is the hadronic distribution 
amplitude $\phi(u,Q^2)$. It describes the probability amplitude to find a 
quark and antiquark of the respective lowest-order Fock state 
carrying  fractional momenta $u$ and $1-u$. For hadrons it is related to the
light-cone wave function of the respective Fock state $\psi$ through \cite{bl}:

\begin{equation}
\phi_{q\bar{q}/\pi}(u,Q^2) ~\sim ~\int_0^{Q^2} \psi_{q\bar{q}/\pi}
(u,\tilde{k_{\perp}}) d^2\tilde{k_{\perp}}
\label{phihad}
\end{equation}
\begin{equation}
Q^2 ~= ~\frac{k_{\perp}^2}{u(1 - u)}
\end{equation}



\section{The Pion Light-Cone Wave Function}

\begin{figure}[htb]
\vspace*{-1cm}
\centerline{\epsfxsize=8cm \epsfbox{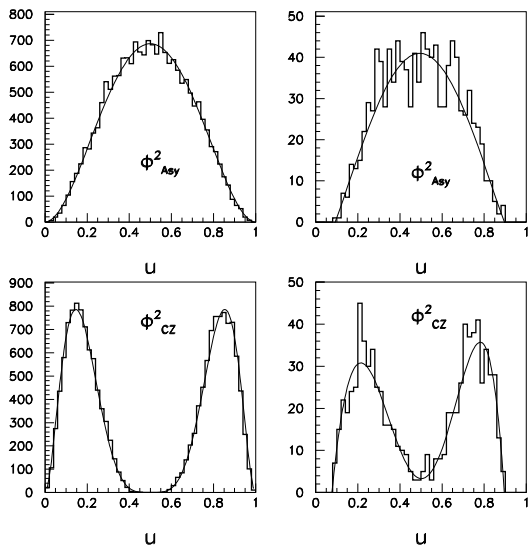}}
\vglue -0.5cm
\caption{Monte Carlo simulations of squares of the two wave functions at the
quark  level (left) and of the reconstructed distributions of di-jets as
detected  (right). }
\label{fig:x_mc}
\end{figure}

For many years two functions were considered to describe the momentum
distribution amplitude of the quark and antiquark in the $|q\bar
{q}\rangle$ configuration. The asymptotic function was calculated  using
perturbative QCD
(pQCD) methods \cite{bl,er,bbgg}, and is the solution to the pQCD
evolution equation for very large $Q^2$ ($Q^2 \rightarrow \infty$):
\begin{equation}
\phi_{Asy}(u) =\sqrt{3} u(1-u).
\label{asy}
\end{equation}
Using QCD sum rules, Chernyak and Zhitnitsky (CZ)
proposed \cite{cz} a function that is expected to be correct for low
Q$^2$:
\begin{equation}
\phi_{CZ}(u) =5\sqrt{3} u(1-u)(1-2u)^2.
\label{cz}
\end{equation}
As can be seen from Eqns. \ref{asy} and \ref{cz} and from Fig.
\ref{fig:x_mc}, there is a large difference between the two functions.
Measurements of the electromagnetic form factors of the pion are related
to the integral over the wave function and the scattering matrix element
and their sensitivity to the shape of the wave function is low
\cite{stesto}. Other open questions are what
can be considered to be high enough $Q^2$ to qualify for perturbative QCD
calculations, what is low enough to qualify for a treatment
based on QCD sum rules, and how to handle the evolution from low to
high $Q^2$.

The concept of the measurement presented here is the following: a high
energy pion
dissociates diffractively on a heavy nuclear target. The first (valence) 
Fock component dominates at large $Q^2$; the other terms are suppressed 
by powers of $1/Q^2$ for each additional parton, according to counting 
rules \cite{stesto,bf}. This is a coherent process in which the quark and 
antiquark break apart and hadronize into two jets. If in this fragmentation 
process the quark momentum is transferred to the jet, measurement of the 
jet momentum gives the quark (and antiquark) momentum. Thus:
$u_{measured} = \frac {p_{jet1}} {p_{jet1}+p_{jet2}}.$
 It has been shown
by Frankfurt {\em et al.} \cite{fms} that the cross section for this
process is prportional to $\phi^2$.

An important assumption is that quark momenta are not modified by nuclear
interactions; i.e., that color transparency \cite{bbgg} is satisfied.
The valence Fock state is far off-shell with mass of several GeV/c$^2$.
This corresponds to a transverse size of about 0.1 fm., much smaller than
the regular pion size. Under these conditions, the valence quark
configuration
will penetrate the target nucleus freely and exhibit the phenomenon of
color transparency. The lifetime of the configuration is given by
$2P_{lab}/(M_n^2 - M_{\pi}^2)$ so at high beam momenta the configuration
is kept small throughout its journey through the nucleus. 
Bertsch {\em et al.} \cite{bbgg} proposed that the small  $|q\bar {q}\rangle$
component will be filtered by the nucleus. Frankfurt {\em et al.} \cite{fms}
show that for $k_t > 1.5 GeV/c$ the interaction with the nucleus is
completely coherent and $ \sigma ( |q\bar {q}\rangle N \rightarrow$di-jets $N$)
is small. This leads to  an $ A^2 $ dependence of the forward amplitude squared.
When integrated over transverse momentum the signature for color transparency 
is a cross section dependence of $A^{4/3}$.

The basic assumption that the momentum carried by the dissociating
$q \bar q$ is transferred to the di-jets was examined by Monte Carlo
(MC) simulations of the asymptotic and CZ wave functions (squared).
The MC samples were allowed to hadronize through the LUND PYTHIA-JETSET
model \cite{mc} and then passed through simulation of the experimental
apparatus. In Fig. \ref{fig:x_mc}, the initial 
distributions at the quark level are compared with the final distributions 
of the detected di-jets. As can be seen, the qualitative features of the 
distributions are retained. 


\section{Experimental Results}

\begin{figure}[htb]
\vspace*{-1cm}
\centerline{\epsfxsize=8cm \epsfbox{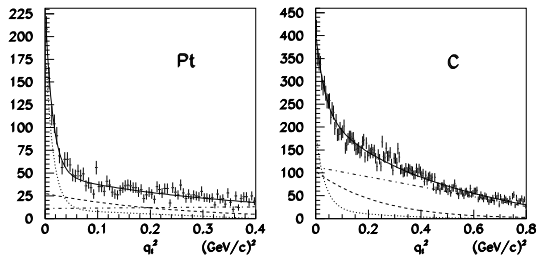}}
\vglue -4.0cm
\caption{ $q_t^2$ distributions of di-jets with $1.5 \leq k_t \leq 2.0$
GeV/c for the platinum and carbon targets. The lines are fits of the MC
simulations to the data: coherent dissociation (dotted line), incoherent
dissociation (dashed line), background (dashed-dotted line), and total fit
(solid line).}
\label{difr_a}
\end{figure}

Results of experimental studies of the pion light-cone wave function
were recently published by the Fermilab E791 collaboration \cite{wfct}.
In the experiment, diffractive dissociation of 500 GeV/c negative pions 
interacting with carbon and platinum targets was measured. 
Diffractive di-jets were required to carry the full beam momentum.
They were identified through the $e^{-bq_t^2}$ dependence
of their yield ($q_t^2$ is the square of the transverse momentum transferred
to the nucleus and $b = \frac{<R^2>}{3}$ where R is the nuclear radius).
Fig. \ref{difr_a} shows the $q_t^2$ distributions
of di-jet events from platinum and carbon. The different
slopes in the low $q_t^2$ coherent region reflect the different
nuclear radii. Events in this region come from diffractive dissociation
of the pion.

\begin{figure}[h]
\centerline{\epsfxsize=8cm \epsfbox{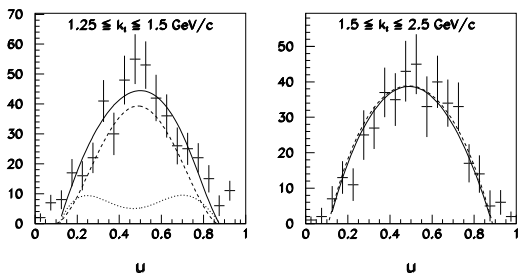}}
\vglue -4.0cm
\caption{ The $u$ distribution of diffractive di-jets from
the platinum target for $1.25 \leq k_t \leq 1.5$ GeV/c (left) and for
$1.5 \leq k_t \leq 2.5$ GeV/c (right). The solid line is a fit to a
combination of the asymptotic and CZ wave functions. The dashed line shows
the contribution from the asymptotic function and the dotted line that of
the CZ function.}
\label{xdatadif}
\end{figure}

For measurement of the wave function the most forward events
($q_t^2 ~< 0.015$ GeV/c$^2$) from the platinum target were used.
For these events, the value of $u$ was computed from the measured 
longitudinal momentaof the jets.  
The analysis was carried out in two windows of $k_t$:
$1.25 ~\rm{GeV/c} ~\leq ~k_t ~\leq ~1.5 ~\rm{GeV/c}$ and
$1.5 ~\rm{GeV/c} ~\leq  ~k_t ~\leq ~2.5 ~\rm{GeV/c}$.
The resulting $u$ distributions are shown in Fig. \ref{xdatadif}.

In order to get a measure of the correspondence between the experimental
results and the calculated light-cone wave functions, the results were
fit with a linear combination of squares of the two wave functions
(right side of Fig. \ref{fig:x_mc}). This
assumes an incoherent combination of the two wave functions and that the
evolution of the CZ function is slow (as stated in \cite{cz}).
The results for the higher $k_t$ window show that
the asymptotic wave function describes the data very well.
Hence, for $k_t > $1.5 GeV/c, which translates
to $Q^2 \sim 10~{\rm (GeV/c)^2}$, the pQCD approach that
led to construction of the asymptotic wave function is reasonable.
The distribution in the lower window is consistent with a significant
contribution from the CZ wave function or may indicate
contributions due to other non-perturbative effects. As the measurements
are done within $k_t$ windows, the results actually represent the square
of the light-cone wave function averaged over $k_t$ in the window: 
$\psi^2_{q\bar{q}}(x,\overline{k_{\perp}})$.

\begin{figure}[htb]
\centerline{\epsfxsize=8cm \epsfbox{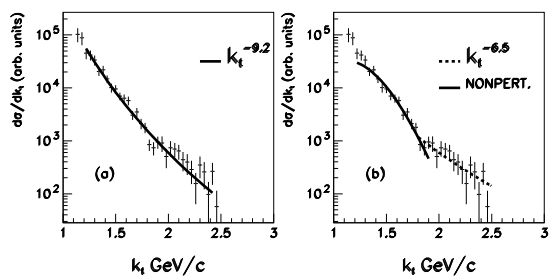}}
\vglue -4.0cm
\caption{ Comparison of the $k_t$ distribution of
acceptance-corrected data with fits to
cross section dependence (a) according to a power law, (b) based on a
nonperturbative Gaussian wave function for low $k_t$ and a power
law for high $k_t$.}
\label{split}
\end{figure}

The $k_t$ dependence of diffractive di-jets is another observable that can
show how well the perturbative calculations describe the data. As shown in
\cite{fms} it is expected to be: ${d\sigma\over dk_t} ~\sim ~k_t^{-6}$.
The results, shown in Fig. \ref{split}, are consistent with this
dependence
only in the region above $k_t \sim$ 1.8 GeV/c, in agreement with the
conclusions from the $u$-distributions. For lower $k_t$ ($Q^2$) values,
non-perturbative effects are expected to be significant.

Finally, verification that the $|q \bar{q} \rangle$ configuration is small
and does not suffer from final state nuclear interaction was done by
observing the color transparency effect for the diffractive dijets.
The A-dependence of the diffractive di-jet yield was measured and found
to have $\sigma \propto A^{\alpha}$ with $\alpha \sim$ 1.5, consistent
with the expected color transparency signal \cite{fms}.


\section{The Photon Light-Cone Wave Function}
\label{sec:phwf}
The photon light-cone wave function can be described in a similar way except
that it has two major components: the electromagnetic and the hadronic
states:
\begin{eqnarray}
\psi_{\gamma} = a |\gamma_p \rangle + b |l^+ l^-\rangle + c |l^+ l^-
\gamma \rangle + (other ~e.m.) \nonumber \\
   + d |q\bar {q}\rangle + e |q\bar
{q}g\rangle + (other ~hadronic) + ... \ .
\label{phwf}
\end{eqnarray}
where $|\gamma_p \rangle$ describes the point bare-photon and  
$|l^+ l^-\rangle$ stands for $|e^+ e^-\rangle,~|\mu^+ \mu^-\rangle$ etc.
Each of these states is a sum over the relevant helicity components.
The wave function 
is very rich: it can be studied for real photons, for virtual
photons of various virtualities, for transverse and longitudinal photons
and the hadronic component may be decomposed
according to the quark's flavor. 
An experimental program based on diffractive dissociation of real and
virtual photons into di-leptons or di-jets is presently on-going at HERA.

I would like to thank Drs. S. Brodsky and L. Frankfurt for many helpful
discussions. I also want to acknowledge the efforts of the E791 collaboration, 
of which I am a member, and particularly my graduate student, R. Weiss-Babai,
for the data presented in this work.

\end{document}